\documentclass[trackchanges,twocolumn,tighten]{aastex63}

\newcommand{\starcluster}{H72.97$-$69.39}

\newcommand{\ltsima}{$\; \buildrel < \over \sim \;$}
\newcommand{\simlt}{\lower.5ex\hbox{\ltsima}}

\newcommand{\ls}{{_<\atop^{\sim}}}

\newcommand{\gs}{{_>\atop^{\sim}}}

\def\arcdeg{\hbox{$^\circ$}}
\def\arcmin{\hbox{$^\prime$}}
\def\arcsec{\hbox{$^{\prime\prime}$}}

\usepackage{amsmath}
\usepackage{graphicx}

\shorttitle{Detection of Diffuse Hot Gas in Young SSC}
\shortauthors{Webb \& Rodriguez et al.}

\begin{document}

\title{Detection of Diffuse Hot Gas Around the Young, Potential Superstar Cluster \starcluster}

\correspondingauthor{Trinity Webb}
\email{webb.916@osu.edu}

\author[0009-0005-7228-9290]{Trinity L. Webb} 
\affil{Department of Astronomy, The Ohio State University, 140 W. 18th Ave., Columbus, OH 43210, USA}

\author[0000-0003-1560-001X]{Jennifer A. Rodriguez}
\affil{Department of Astronomy, The Ohio State University, 140 W. 18th Ave., Columbus, OH 43210, USA}

\collaboration{8}{These authors contributed equally to this work.}

\author[0000-0002-1790-3148]{Laura A. Lopez}
\affil{Department of Astronomy, The Ohio State University, 140 W. 18th Ave., Columbus, OH 43210, USA}
\affil{Center for Cosmology and AstroParticle Physics, The Ohio State University, 191 W. Woodruff Ave., Columbus, OH 43210, USA}
\affil{Center for Computational Astrophysics, Flatiron Institute, 162 5th Avenue, New York, NY 10010, USA}

\author[0000-0003-4423-0660]{Anna L. Rosen}
\affil{Department of Astronomy, San Diego State University, 5500 Campanile Dr, San Diego, CA 92182, USA}
\affil{Computational Science Research Center, San Diego State University, 5500 Campanile Dr, San Diego, CA 92182, USA}

\author[0000-0002-0041-4356]{Lachlan Lancaster}
\thanks{Simons Fellow}
\affil{Department of Astronomy, Columbia University,  550 W 120th St, New York, NY 10025, USA}
\affil{Center for Computational Astrophysics, Flatiron Institute, 162 5th Avenue, New York, NY 10010, USA}

\author[0000-0001-6576-6339]{Omnarayani Nayak}
\affil{Space Telescope Science Institute, 3700 San Martin Drive, Baltimore, MD 21218, USA}

\author[0000-0002-5456-523X]{Anna F. McLeod}
\affil{Centre for Extragalactic Astronomy, Department of Physics, Durham University, South Road, Durham DH1 3LE, UK}
\affil{Institute for Computational Cosmology, Department of Physics, University of Durham, South Road, Durham DH1 3LE, UK}

\author[0009-0003-6803-2420]{Paarmita Pandey}
\affil{Department of Astronomy, The Ohio State University, 140 W. 18th Ave., Columbus, OH 43210, USA}
\affil{Center for Cosmology and AstroParticle Physics, The Ohio State University, 191 W. Woodruff Ave., Columbus, OH 43210, USA}

\author[0000-0002-4606-4240]{Grace M. Olivier}
\affil{Department of Physics and Astronomy and George P. and Cynthia Woods Mitchell Institute for Fundamental Physics and Astronomy, Texas A\&M Univeersity, 4242 TAMU, College Station, TX 77843-4242 USA}

\begin{abstract}

We present the first Chandra X-ray observations of \starcluster, a highly-embedded, potential super-star cluster (SSC) in its infancy located in the star-forming complex N79 of the Large Magellanic Cloud. We detect particularly hard, diffuse X-ray emission that is coincident with the young stellar objects (YSOs)  
identified with JWST, and the hot gas fills cavities in the dense gas mapped by ALMA. The X-ray spectra are best fit with either a thermal plasma or power-law model, and assuming the former, we show that the X-ray luminosity of $L_{\rm X} = (1.0\pm0.3)\times10^{34}~{\rm erg}~{\rm s}^{-1}$ is a factor of $\sim$20 below the expectation for a fully-confined wind bubble. Our results suggest that stellar wind feedback produces diffuse hot gas in the earliest stages of massive star cluster formation and that wind energy can be lost quickly via either turbulent mixing followed by radiative cooling or by physical leakage. 

\end{abstract}


\section{Introduction} 
\label{sec:intro}

Massive stars are born in clustered environments \citep{Krumholz19}, depositing substantial energy and momentum to the surrounding interstellar medium (ISM) through a variety of feedback mechanisms. In particular, fast, line-driven stellar winds (with velocities of $v_{\rm w} \sim 10^{3}~{\rm km}~{\rm s}^{-1}$) sweep up surrounding gas and create low-density cavities shock-heated to $\sim10^{7}$~K temperatures \citep{Castor75,Weaver77,Canto00,Stevens03,HarperClark09}. Diffuse X-ray emission associated with these fast stellar winds has been detected from numerous massive star clusters (MSCs) in the Milky Way \citep{Moffat02,YZ02,Townsley03,Muno06,Townsley11} and the Magellanic Clouds \citep{Townsley06,Lopez11,Lopez14}. Although the integrated kinetic energy carried in the stellar winds is comparable to the kinetic energy delivered by supernova explosions \citep{Agertz13}, the actual dynamical impact of the stellar wind feedback and how it evolves over time remains uncertain.

\begin{figure*}[t]
    \centering    
    \includegraphics[width=0.45\textwidth]{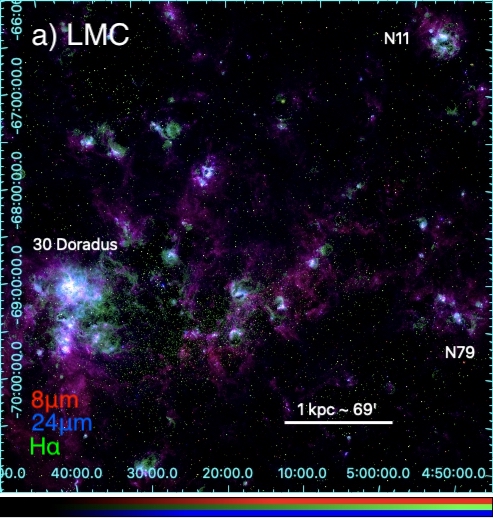}  \includegraphics[width=0.45\textwidth]{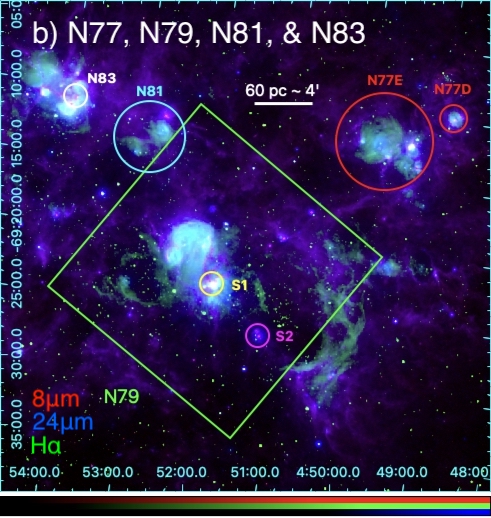} \\
   \includegraphics[width=0.45\textwidth]{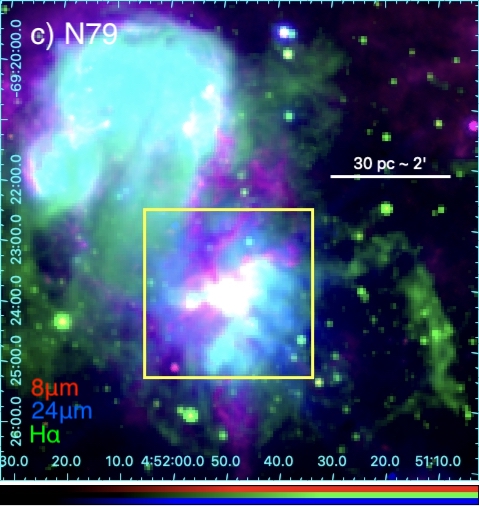}
   \includegraphics[width=0.45\textwidth]{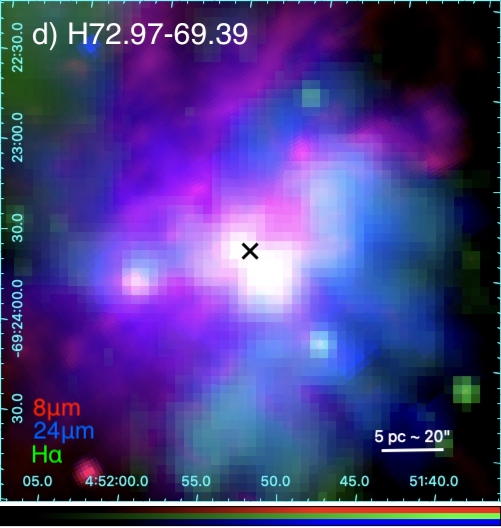}
    \caption{Three-color images, with Spitzer SAGE 8$\mu$m in red, Spitzer SAGE 24$\mu$m in blue \citep{Meixner06}, and H$\alpha$ in green \citep{Smith98}. North is up, East is left. (a) Large-scale, three-color image of the Large Magellanic Cloud. (b) Zoomed in three-color image of N77, N79, N81, and N83 identified by \cite{Henize56}. N79, N83, and N77 have also been referred to as N79 South, N79 East, and N79 West in \cite{Ochsendorf17}, respectively. Two Spitzer-identified clusters of young stellar objects (YSOs) within N79 are circled in yellow (the location of \starcluster) and magenta, known as N79S1 and N79S2, respectively \citep{Nayak24}. The green box surrounding N79 is the field of view of the Chandra ACIS-I observations. (c) Zoomed in three-color image of N79. The yellow box indicates the area surrounding H72.97-69.39 as shown in panel d. (d) Zoomed in three-color image in the vicinity of \starcluster\ (black X).}
    \label{fig:LMC-H72}
\end{figure*}

One open issue is the role of winds in the early evolution of MSCs, and the study of young, embedded sources can provide important insights. Toward this end, the highly embedded N79 star-forming complex in the southwestern region of the Large Magellanic Cloud (LMC; see Figure~\ref{fig:LMC-H72}), which has a star-formation efficiency of $\sim$2 times that of the starburst region 30 Doradus (hereafter 30 Dor; \citealt{Ochsendorf17}), is an ideal target to study the role of stellar wind feedback at the onset of star cluster formation. The N79 region has several CO sub-complexes spanning $\sim$500 pc (see Figure~\ref{fig:LMC-H72}; \citealt{Wong11,Ochsendorf17}) and harbors a newly forming ($<$0.5 Myr old; \citealt{Ochsendorf17}), potential proto-superstar cluster (SSC) \starcluster \footnote{This source is catalogued as HSOBMHERICC J72.971176$-$69.391112 \citep[][; hereafter \starcluster]{Seale14} and will be referred to as \starcluster\ throughout this paper. The position of \starcluster\ was refined by \cite{Nayak19} based on the compact core's continuum emission to R.A. 72.972201 and decl. $-$69.391301.} based on its inferred bolometric luminosity ($\sim 2 \times 10^6 L_{\sun}$) and accelerating star-formation rate \citep{Seale14,Ochsendorf17,Nayak19}.  In comparison, R136 (the star cluster powering 30 Dor) has a luminosity of $\sim 7.8 \times 10^7 L_{\sun}$ \citep{Malumuth94} and a decelerating star-formation rate. Both 30 Dor and N79 are on opposite leading edges of the LMC's spiral arms \citep{Ochsendorf17}, possibly facilitating their starburst activity due to large-scale dynamical inflows along the tidal tails. While R136 is 2~Myr old \citep{Hunter95}, \starcluster\ is in the earliest stages of formation, offering an interesting point of comparison.

Previous studies of \starcluster\ have mapped the N79 star-forming complex in optical, infrared, and sub-mm wavelengths \citep{Ochsendorf17, Nayak19,Nayak24}, tracing the young stars, the cold molecular gas, and the dust. Here we study the X-ray emission of \starcluster\ with the Chandra X-ray Observatory, and we explore stellar-wind feedback at an early stage ($\ls$0.5 Myr) in star formation. Specifically, we show the spatial distribution of the hot gas relative to the other gas phases, dust, and stars as well as compare the X-ray luminosity and hot-gas temperature predictions from wind bubble models. For comparison, the sample of 32 H{\sc ii} regions (including the full N79-South complex) examined by \cite{Lopez14} had estimated ages of 3 $-$ 10 Myr, when the sources' shells had expanded $\sim$4 $-$ 150~pc.  The other SSC in the LMC, R136 in 30~Dor, has been observed extensively by Chandra, including an X-ray Visionary Project (PI: L. Townsley) totaling $\sim$2~Ms \citep{Townsley24}. Chandra studies of 30~Dor have found diffuse, hot gas filling the five H$\alpha$ shells of the region, producing a total X-ray luminosity of $L_{\rm X} = 4.5\times10^{36}~{\rm erg}~{\rm s}^{-1}$ from the diffuse emission in the 0.2$-$2.0~keV band \citep{Townsley06,Lopez11}. 

This paper is structured as follows. In Section~\ref{sec:data}, we describe the new Chandra observations of \starcluster\ and discuss the analysis to produce the X-ray images and spectra. In Section~\ref{sec:results}, we present the results, including the spatial extent, the nature, and the association of the X-ray emission with young stellar objects (YSOs) and anti-coincidence with the dense gas. In  Section~\ref{subsec:wind_xray_predict}, we interpret the results in the context of wind bubble models, and in Section~\ref{subsec:YSO}, we compare the X-ray emission of \starcluster\ to other Milky Way and LMC sources (by comparison, most SMC H{\sc ii} regions are not X-ray detected; \citealt{Lopez14}). We conclude in Section~\ref{sec:conc}.

\section{Data Analysis} \label{sec:data}

\subsection{Imaging} \label{sec:images}

\begin{figure*}
    \centering    \includegraphics[width=0.45\textwidth]{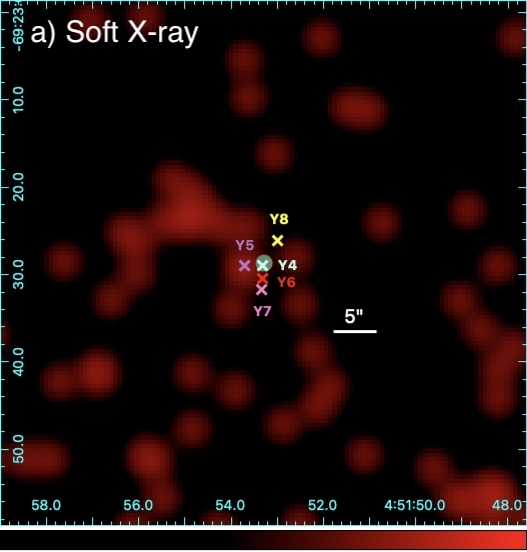}
   \includegraphics[width=0.45\textwidth]{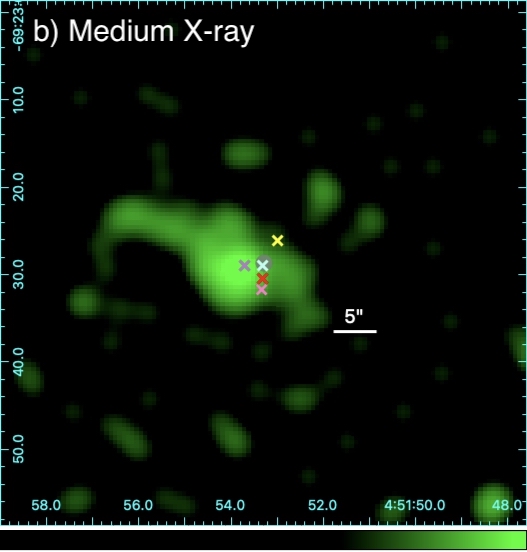} \\
\includegraphics[width=0.45\textwidth]{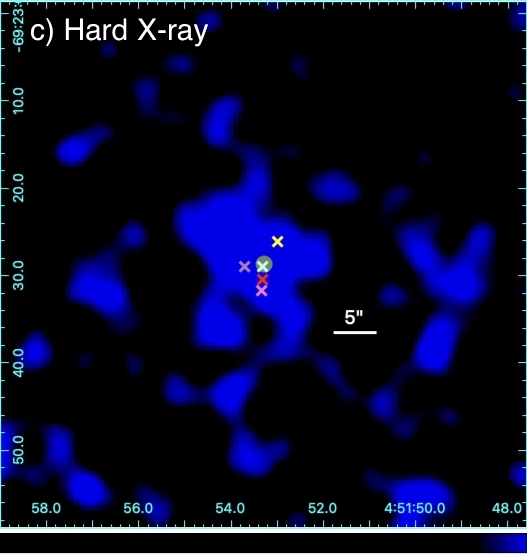}
  \includegraphics[width=0.45\textwidth]{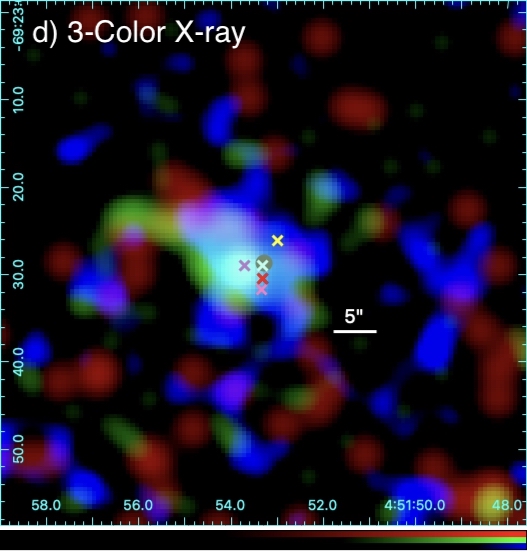}
    \caption{X-ray images of \starcluster\ in soft [$0.5-1.2$~keV] (a), medium [$1.2-2.0$~keV] (b), hard [$2.0-7.0$~keV](c), and broad band [$0.5-7.0$~keV] (d). The scale bar is 5\arcsec\ $\approx$ 1.2~pc, and the images are 1\arcmin$\times$1\arcmin\ in size. A gray circle marks the position of \starcluster, and five YSOs identified by \cite{Nayak24} with JWST MRS data are labeled with X symbols. North is up, East is left. The X-ray emission is extended $\approx$10\arcsec\ and peaks $\approx$5\arcsec\ offset from the position of \starcluster.}
    \label{fig:3colorxray_panel}
\end{figure*}

\begin{figure*}
    \centering    \includegraphics[width=.9\textwidth]{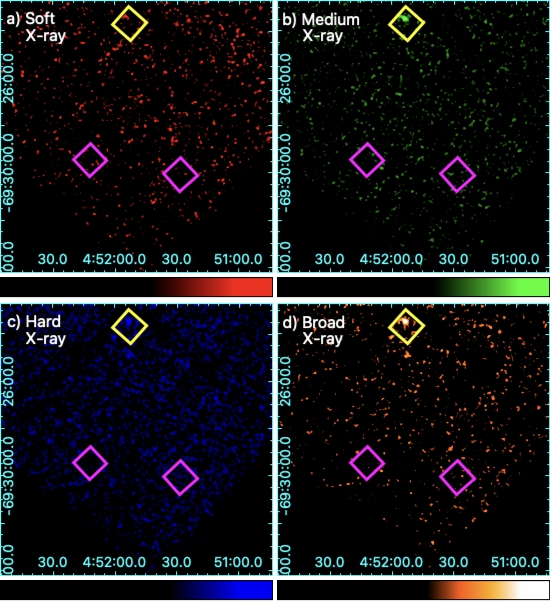}
    \caption{Zoomed in Chandra ACIS-I images showing the source region (yellow box) and background regions (magenta boxes) used to create spectra of \starcluster. Each region is 1\arcmin$\times$1\arcmin. The distance from the center of the source region to the center of the left-hand background region is $\approx6.0$\arcmin\ ($\approx 87.5$~pc), and the distance to the right-hand background region is $\approx6.7$\arcmin\ ($\approx 98.1$~pc). Panel (a) shows the soft [$0.5-1.2$~keV] X-ray band in red, panel (b) shows the medium [$1.2-2.0$~keV] in green, panel (c) shows the hard [$2.0-7.0$~keV] in blue, and panel (d) shows the broad-band [$0.5-7.0$~keV] in orange.}
    \label{fig:regions}
\end{figure*}

\starcluster\ in N79 was observed three times with Chandra ACIS-I in July 2021 for a total of 98~ks (ObsIDs 22473 [29~ks], 23062 [39~ks], and 25091 [30~ks]). These Chandra datasets, obtained by the Chandra X-ray Observatory, are contained in~\dataset[doi:10.25574/cdc.292] {https://doi.org/10.25574/cdc.292}. Data were reprocessed and reduced using Chandra Interactive Analysis of Observations {\sc ciao} version 4.15 \citep{CIAO2006}. We used the \textit{merge\_obs} function to produce exposure-corrected images (images that are normalized by the effective area across the detectors) in the soft (0.5$-$1.2 keV), medium (1.2$-$2.0 keV), hard (2.0$-$7.0 keV), and broad (0.5$-$7.0 keV) energy bands (see Figure~\ref{fig:3colorxray_panel}). We detect 156$\pm$28 net, background-subtracted, broad-band counts within a 1\arcmin$\times$1\arcmin\ area surrounding \starcluster\ in the merged, broad-band image. Background subtraction was performed using the regions shown in Figure~\ref{fig:regions}, and the total source and background counts in each band are listed in Table~\ref{table:counts}. The source region is not statistically significantly detected above the background in the soft (0.5$-$1.2~keV) band, and the signal is strongest in the hard (2.0$-$7.0~keV) X-ray band.

\begin{deluxetable}{lccr}
\tablecolumns{4}
\tablewidth{0pt} \tablecaption{Detected Counts Per Band in Source and Background Regions\tablenotemark{a} \label{table:counts}} 
\tablehead{\colhead{Band} & \colhead{Source} & \colhead{Background} & \colhead{Net} \\
\colhead{} & \colhead{Counts} & \colhead{Counts} & \colhead{Counts}}
\startdata
Broad (0.5$-$7.0~keV) & 558$\pm$24 & 805$\pm$28 & 156$\pm$28 \\
Soft (0.5$-$1.2~keV) & 75$\pm$9 & 147$\pm$12 & 2$^{+11}_{-2}$ \\
Medium (1.2$-$2.0~keV) & 128$\pm$11 & 147$\pm$12 & 54$\pm$13 \\
Hard (2.0$-$7.0~keV) & 355$\pm$19 & 548$\pm$23 & 81$\pm$22 \\
\enddata
\tablenotetext{a}{The source and background regions have areas of 14875 pixel$^2$ ($\sim$3600 arcsec$^2$) and 29742 pixel$^2$ ($\sim$7200 arcsec$^2$), respectively.}
\end{deluxetable}

\subsection{Spectroscopy}

In addition to the imaging analysis, we conduct a spectral analysis of the X-rays from \starcluster. Using the {\sc ciao} command {\it specextract}, we extracted spectra from a source region with an area of 1\arcmin$\times$1\arcmin (see Figure~\ref{fig:regions}). The background spectra were extracted from two 1\arcmin$\times$1\arcmin~regions to the south $\sim$6\arcmin\ of the source region. Background regions were selected to avoid X-ray point sources identified with the {\sc ciao} command {\it wavdetect} and to not coincide with any bright 24 $\mu$m emission. 

We jointly fit the spectra using either thermal or non-thermal plasma models in XSPEC version 12.13.0c \citep{XSPEC}. The best fit parameters for these models can be seen in Table~\ref{table:spectralresults}. The thermal plasma model included a multiplicative constant (\textsc{const}), two absorption components (\textsc{phabs*vphabs}), and one additive thermal plasma component (\textsc{apec}). The \textsc{const} component was allowed to vary to account for the slight variations in flux between the three observations. The \textsc{phabs} component is a photoelectric absorption of $M(E) = \rm{exp}[-\eta_{\rm H} \sigma(E)]$, where $\sigma(E)$ is the photo-electric cross-section and $\eta_{\rm H}(E)$ is the equivalent hydrogen column\footnote{See heasarc.gsfc.nasa.gov/xanadu/xspec/manual/XSmodelPhabs.html}. We fixed the first \textsc{phabs} to account for the column density of the Milky Way and set it to the neutral atomic hydrogen (H{\sc i}) density observed in the direction of N79, $N_{\rm H}^{\rm MW}=3.24\times 10^{21}\: \mathrm{cm^{-2}}$ \citep{HI}. By adopting this value, we neglect the contribution from the molecular and ionized components to the hydrogen column density. We note that \cite{wilms00} found that atomic hydrogen is responsible for $\sim$80\% of the total hydrogen column density, and our selected $N_{\rm H}^{\rm MW}$ is comparable to the total hydrogen column density of $N_{\rm H,tot} = 3.42\times10^{21}$~cm$^{-2}$ derived from \cite{Willingale13}, assuming uniform elemental abundances and $N_{\rm H,tot} = N_{\rm HI} + 2 N_{\rm H_{2}}$, with the atomic hydrogen column density $N_{\rm HI}$ derived from 21-cm radio observations and the molecular hydrogen column density $N_{\rm H_{2}}$ from X-ray afterglows of gamma-ray bursts (GRBs).

The \textsc{vphabs} component represented the column density of the LMC toward \starcluster\ and was set to $N_{\rm H}^{\rm LMC}=5.5\times 10^{21}\: \mathrm{cm^{-2}}$, the upper-limit value defined by the XSPEC \textsc{error} command when $N_{\rm H}^{\rm LMC}$ was allowed to vary. The \textsc{apec} component, reflecting an optically thin thermal plasma, is defined by its temperature $kT$, metal abundances, redshift, and a normalization factor. Solar abundances were adopted from \cite{Asplund09}. The abund command in XSPEC\footnote{https://heasarc.gsfc.nasa.gov/xanadu/xspec/\\ manual/node116.html} was used to set the abundance to 0.5 solar metallicity in the \textsc{apec} and \textsc{vphabs} components to match the values of the LMC ISM \citep{kurt98,maggi16}. The temperature and normalization were allowed to vary in the fits. 

For comparison, we also tried fitting the data with a power-law component (\textsc{powerlaw}) representing non-thermal emission in the place of the \textsc{apec} component. Given the low signal (156 net broad-band counts), we did not attempt a spectral model that included both thermal and non-thermal components as it would be under-constraining (i.e., too many free parameters for the limited degrees of freedom).

\section{Results} \label{sec:results}

Figure~\ref{fig:3colorxray_panel} shows the exposure-corrected, three-color X-ray image of the soft ($0.5-1.2$~keV), medium ($1.2-2.0$~keV), and hard ($2.0-7.0$~keV) X-ray bands around \starcluster. The X-rays are spatially extended $\sim$10\arcsec\ $\approx$ 2.4~pc in radius, much greater than the $\approx$0.5\arcsec\ on-axis point spread function (PSF) of Chandra ACIS\footnote{https://cxc.harvard.edu/proposer/POG/html/chap6.html}. The peak of the X-rays is spatially offset $\approx$5\arcsec\ $\approx$1.2~pc east of the star cluster. It is possible that some emission arises from unresolved point sources coincident with the diffusion emission, but no resolved point sources are apparent in the vicinity of \starcluster.

The X-rays (particularly the medium and hard bands) are spatially coincident with five YSOs (identified as Y4, Y5, Y6, Y7, and Y8) identified with recent JWST MRS observations \citep{Nayak24}, as marked in Figure~\ref{fig:3colorxray_panel}. Y4 is coincident with \starcluster, and it is the only one of the five YSOs without emission lines associated with polycyclic aromatic hydrocarbons (PAHs) in the MRS spectra, likely stemming from ionizing radiation destroying surrounding PAHs (e.g., \citealt{madden06,gordon08,montillaud13,egorov23}). Y5 is located at the peak of the broad-band X-ray emission and is the only YSO that coincides with soft X-rays. All five YSOs coincide with hard X-ray emission. While their physical association with the hard X-rays is not entirely certain, it is possible that they arise from colliding stellar winds in binary systems \citep{Rauw16}. All except Y8 have coincident medium X-ray emission. 

\begin{figure*}
    \centering
    \includegraphics[width=\textwidth]{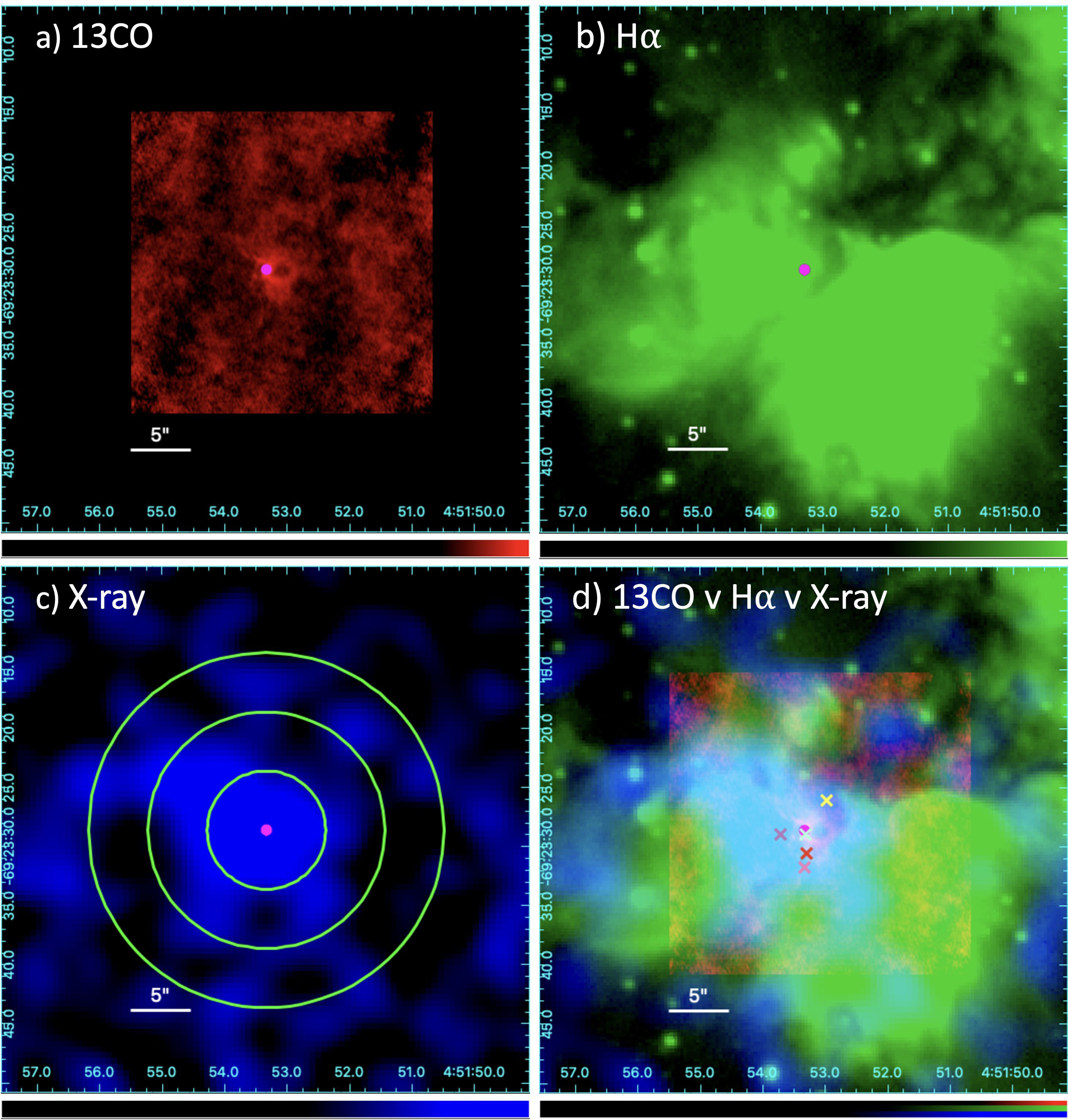}
    \caption{Comparison of the (a) $^{13}$CO emission observed with ALMA in red by \cite{Nayak19} with the (b) DeMCELS H$\alpha$ emission from \cite{Points24} in green and (c) Chandra broad-band ($0.5-7.0$~keV) X-ray emission in blue. The five YSOs identified with JWST by \cite{Nayak24} are marked with Xs in panel (d). \starcluster\ is denoted by a magenta circle in all panels. The diffuse X-rays appear to fill the low-density cavities where $^{13}$CO is dim. Annuli in panel (c) were used to produce the multi-band surface-brightness profiles in Figure~\ref{fig:Profile_panel}. All images are 0.75\arcmin$\times$0.75\arcmin\ in size, and the scale bar is 5\arcsec $\approx$~1.2 pc at the distance of the LMC.}
    \label{fig:CO13vXray}
\end{figure*}

\begin{figure}
    \centering
    \includegraphics[width=0.8\columnwidth]{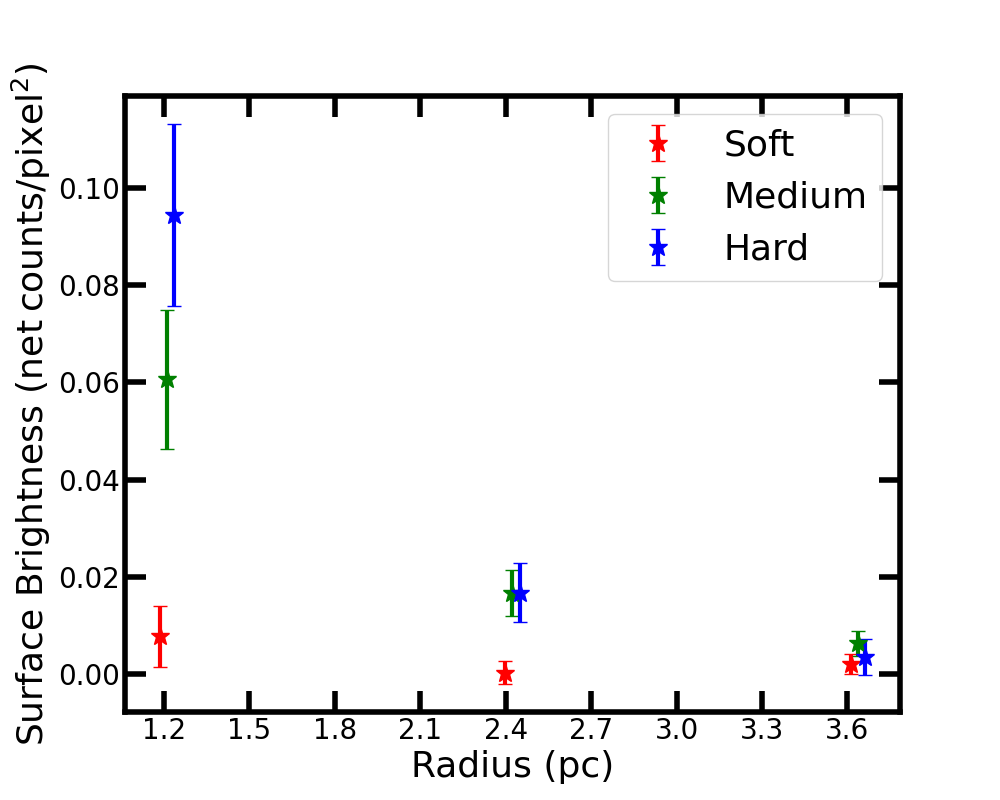}
    \caption{Background-subtracted, surface-brightness profile (in units of net counts pixel$^{-2}$) of the soft (red), medium (green), and hard (blue) X-ray bands for the three annuli in Figure~\ref{fig:CO13vXray}a. The hard X-rays have the greatest surface brightness near \starcluster, and the medium and hard X-rays have comparable surface brightnesses at radii $\gs$2.4~pc. The soft X-rays are not statistically significantly detected in any annulus.}
    \label{fig:Profile_panel}
\end{figure}

To quantify the spatial distribution of the X-ray emission, we measured the background-subtracted surface brightness of the soft, medium, and hard X-ray bands from three annuli centered on \starcluster\ with radii of 5\arcsec ($\approx$1.2~pc), 10\arcsec ($\approx$2.4~pc), and 15\arcsec\ ($\approx$3.6~pc; see Figure~\ref{fig:CO13vXray}) and produced the profiles plotted in Figure~\ref{fig:Profile_panel}. We find that the hard X-rays have the highest net surface brightness near \starcluster, while the medium and hard X-ray surface brightnesses are comparable at radii of 2.4~pc and 3.6~pc. These profiles suggest that the X-ray emission is hardest around \starcluster\ and softens with distance from the star cluster. X-ray hardness variations across the region may reflect higher column densities $N_{\rm H}^{\rm LMC}$ that are absorbing softer X-rays, elevated hot gas temperatures $kT$ producing more hard X-rays, and/or the presence of colliding-wind massive binaries \citep{Rauw16}. 

In Figure~\ref{fig:CO13vXray}, we compare the broad-band X-ray emission with the $^{13}$CO maps \citep{Nayak19} and the DeMCELS H$\alpha$ data \citep{Points24}. The X-rays appear to fill the low-density cavities where the $^{13}$CO is dim. The H$\alpha$ is coincident with the X-ray emission surrounding \starcluster\ and the YSOs from \cite{Nayak24}, revealing a bubble that is filled with hot gas, producing the X-ray emission.

\begin{figure}
    \centering    
    \includegraphics[width=\columnwidth]{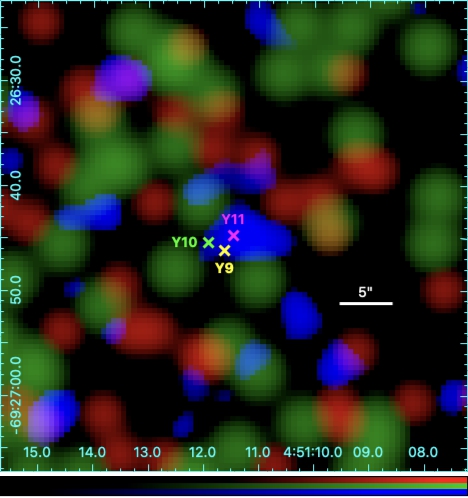}
    \caption{Three-color X-ray image of N79S2, where the soft X-rays (0.5$-$1.2 keV) are in red, medium X-rays (1.2$-$2 keV) are in green, and hard X-rays (2$-$7 keV) are in blue. The three YSOs identified with the JWST MRS data from \cite{Nayak24} are labeled. The hard X-rays dominate this source, likely because of high absorbing column toward it. The image is 0.75\arcmin$\times$0.75\arcmin; the scale bar is 5\arcsec\ $\approx$ 1.2 pc at the distance of the LMC.}
\label{fig:3colorxray_S2panel}
\end{figure}

Although our Chandra program targeted \starcluster, N79S2 (see Figure~\ref{fig:LMC-H72}) -- another part of N79 South observed with JWST MRS by \cite{Nayak24} -- was also in the field of view. In Figure~\ref{fig:3colorxray_S2panel}, we show the three-color X-ray image of N79S2. We detect 12$\pm$5 net, broad-band counts that are coincident with the three YSOs Y9, Y10, and Y11 identified by \cite{Nayak24}. In particular, the hard X-rays are concentrated to the location of Y11. The angular extent of the emission, $\sim$5\arcsec, is consistent with the off-axis PSF of Chandra, and thus, we are unable to distinguish whether the emission is point-like or diffuse in nature.

\begin{figure*}
    \centering    
    \includegraphics[width=\textwidth]{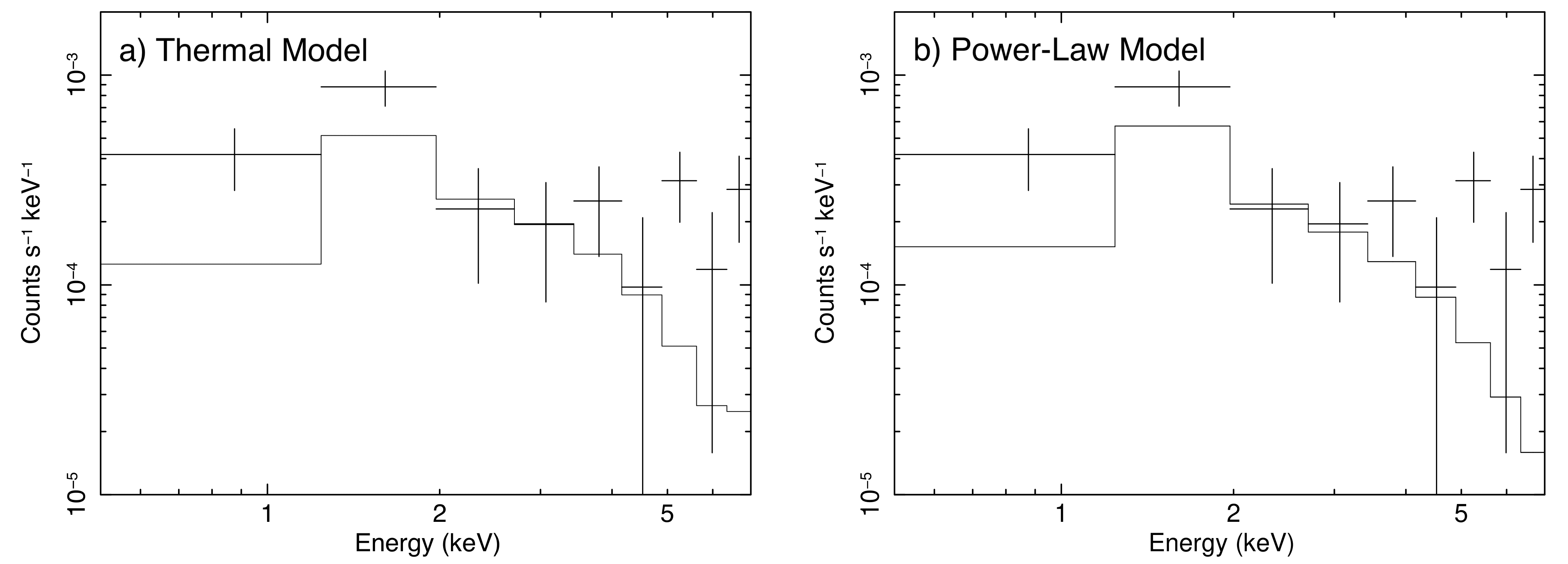}
    \caption{Extracted, background-subtracted spectra from a 1\arcmin$\times$1\arcmin\ region around \starcluster\ with the best-fit thermal plasma model (a) and power-law model (b) overplotted. The best-fit parameters are listed in Table~\ref{table:spectralresults}. Both models produce statistically good fits and predict similar X-ray luminosities from the source.}
    \label{fig:Spectrum_panel}
\end{figure*}

The X-ray spectra from \starcluster\ (shown in Figure~\ref{fig:Spectrum_panel}) give additional constraints on the emission. Both the single thermal plasma model and the power-law model yield statistically good fits (see Table~\ref{table:spectralresults}). We find an upper-limit on the LMC column density toward the source of $N_{\rm H}^{\rm LMC} = 5.5\times10^{21}$~cm$^{-2}$, and we fixed it to this value before estimating the other parameters. The thermal plasma model gave a $\chi ^{2}$/d.o.f. = 54/52 = 1.04 and a best-fit $kT = 3.5^{+32.1}_{-2.2}$~keV. The power-law model produces similar results, with a $\chi ^{2}$/d.o.f. = 53/47 = 1.15 and a best-fit photon index of $\Gamma = 2.2^{+0.3}_{-0.6}$. The resultant luminosities for the two models are comparable and were $L_{\rm X} = (1.0\pm0.3) \times 10^{34} \: \mathrm{erg} \: \mathrm{s}^{-1}$ and $L_{\rm X} = (1.2\pm0.3) \times 10^{34} \: \mathrm{erg} \: \mathrm{s}^{-1}$, respectively. 

\begin{deluxetable*}{lcccccr}
\tablecolumns{7}
\tablewidth{0pt} \tablecaption{Spectral Model Results \label{table:spectralresults}} 
\tablehead{\colhead{Model} & \colhead{$kT$} & \colhead{$\Gamma$} & \colhead{$\chi^2$/d.o.f.} & \colhead{Absorbed Flux\tablenotemark{a}} & \colhead{Emitted Flux\tablenotemark{a}} & \colhead{Luminosity} \\
\colhead{} & \colhead{(keV)} & \colhead{} & \colhead{} & \colhead{(erg~cm$^{-2}$~s$^{-1}$)} & \colhead{(erg~cm$^{-2}$~s$^{-1}$)} & \colhead{(erg~s$^{-1}$)}}  
\startdata
Thermal Plasma & 3.5$^{+32.1}_{-2.2}$ & -- & 54/52 & (2.3$\pm$0.7)$\times10^{-14}$ & (3.4$\pm$1.0)$\times10^{-14}$ & $(1.0\pm0.3) \times 10^{34}$ \\
Power Law & -- & 2.2$^{+0.3}_{-0.6}$ & 53/47 & (2.2$\pm$0.6)$\times10^{-14}$ & (3.9$\pm$1.1)$\times10^{-14}$ & $(1.2\pm0.3) \times 10^{34}$ \\
\enddata
\tablenotetext{a}{Absorbed fluxes are for the 0.5$-$7.0 keV band adopting a LMC and Milky Way column density of $N_{\rm H}^{\rm LMC} = 5.5\times10^{21}$~cm$^{-2}$ and $N_{\rm H} = 3.24\times10^{21}$~cm$^{-2}$, respectively. Emitted fluxes are the unabsorbed fluxes in the same energy bands.}
\end{deluxetable*}

\section{Discussion} \label{sec:disc}

In Section~\ref{sec:results}, we found that \starcluster\ (and the associated YSOs identified with JWST) has coincident diffuse X-ray emission. The flux in the medium and hard bands is more centrally concentrated than the soft X-rays, and the X-ray spectrum can be modeled as a thermal plasma or with a power-law. In Section~\ref{subsec:wind_xray_predict}, we compare these results to predictions from wind models, and in Section~\ref{subsec:YSO}, we discuss how the X-ray properties of \starcluster\ compare to those of other young MSCs in the Milky Way and LMC.

\subsection{Comparison to Wind Models X-ray Predictions}
\label{subsec:wind_xray_predict}

In an idealized H{\sc ii} region powered by a MSC, wind energy is injected at a rate of 

\begin{equation}
\label{eq:wind_luminosity}
L_{\rm w} = \sum_{i=1}^{N} \frac{1}{2} \dot{M}_{{\rm w},i} v_{{\rm w},i}^{2},
\end{equation}

\noindent
where $\dot{M}_{\rm w,i}$ and $v_{\rm w,i}$ are the mass-loss rate and wind velocity for individual stars $i$ that are summed over the total $N$ stars in the cluster. Typical values of $L_{\rm w}$ for Milky Way and LMC MSCs are $L_{\rm w} \sim 10^{37} - 10^{39}$~erg~s$^{-1}$ (e.g., \citealt{Smith06,Rosen14}). We can estimate $L_{\rm w}$ for \starcluster\ assuming $L_{\rm bol} = 2.2\times10^{6}~L_{\sun}$ \citep{Ochsendorf17} and relating it to the expected total wind luminosity using stellar population synthesis and a prescription for wind mass loss and velocity. 

In particular, we first download isochrones of $[{\rm Fe/H}] = -0.25$ (meant to approximate $Z = 0.5 Z_{\odot}$, the LMC ISM metallicity: \citealt{kurt98,maggi16}) from the MIST\footnote{https://waps.cfa.harvard.edu/MIST/} database \citep{Choi16,Dotter16}. We then generate a large (total mass $10^6 M_{\odot}$) sample of masses following the Kroupa IMF \citep{Kroupa01} with minimum and maximum masses of $M_{\rm min} = 0.01 \, M_{\odot}$ and $M_{\rm max} = 200\, M_{\odot}$. We use the MIST isochrones to derive bolometric luminosities for each star in the sample as a function of time. Assuming the mass-loss rates of \cite{Vink00,Vink01}, we also calculate the wind mass loss and wind velocities of all stars with mass $M_* > 8 M_{\odot}$ in the sample and use Equation \ref{eq:wind_luminosity} to calculate their wind luminosities. 

At each point in time (time-steps determined by the age steps of the MIST isochrones), we sum the bolometric and mechanical wind luminosities over all stars in the sample and divide by the total mass of the sample to arrive at population-averaged values for $L_{\rm bol}/M_{*}$ and $L_{\rm w}/M_*$. We average these ratios over ages $<$2~Myr, roughly the period over which wind luminosities remain roughly constant before the onset of SNe \citep{lancaster21a}. We obtain $L_{\rm bol}/M_{*} = 1.37 \times 10^3 \, L_{\odot}/M_{\odot}$ and $L_{\rm w}/M_{*} = 4.20 \, L_{\odot}/M_{\odot}$. Given the $L_{\rm bol}$ of \starcluster, we estimate a mechanical wind luminosity of $L_{\rm w} = 6.7\times 10^3\, L_{\odot} = 2.6\times 10^{37} \, {\rm erg~s}^{-1}$.

In order to relate this value to an expected X-ray luminosity $L_{\rm X}$ in the bubble, we need to consult a model for wind-blown bubble evolution. The theoretical models of \cite{Castor75} and \cite{Weaver77} predict X-ray luminosities assuming the shock-heated gas is confined by a cool shell of swept-up ISM. Additionally, this cool shell is heated by the bubble's hot gas through thermal conduction, resulting in evaporation and mass loading of the bubble interior, resulting in higher X-ray luminosities. An alternative model by \cite{Chevalier85} ignores the surrounding ISM and assumes a steady, freely-expanding wind, equivalent to assuming that the ``free-wind" of the \cite{Weaver77} model occupies the entire bubble volume. As the Castor and Weaver models can over-predict X-ray luminosity and Chevalier \& Clegg can under-predict it, \cite{HarperClark09} introduced an intermediate model, where hot gas expands into an inhomogeneous ISM, creating a porous shell from which hot gas can escape/leak. Leakage is one mechanism by which energy can be lost from bubbles producing lower-than-expected X-ray luminosities (e.g., \citealt{Lopez11,Rosen14}). Other mechanisms have been proposed as possible sinks for the wind energy, chiefly turbulent mixing with the surrounding gas followed by radiative cooling \citep{Rosen14,lancaster21b, Rosen2022}. 

The differences between these wind energy loss channels are likely dependent on the detailed density and temperature structure in the bubble's interior, particularly at its interface. Here we simply present the predictions for X-ray luminosity $L_{\rm X}$ and radius $R$ assuming a single density and temperature as predicted for a \cite{Weaver77} model. The Weaver et al. model requires an assumption of the background number density of the surrounding cloud; measurements from \cite{Ochsendorf17} indicate a molecular gas mass $M_{\rm gas, H_2} \approx 10^5\, M_{\odot}$ within $R = 10^{1.4}\, {\rm pc} \approx 25 \, {\rm pc}$ (see their Figure 5). In the same region atomic gas contributes about one-third as much mass and ionized gas contributes negligibly, indicating an average gas mass density of $\rho \approx 2 \, M_{\odot}\, {\rm pc}^{-3} = 1.4\times10^{-22}$~g~cm$^{-3}$.

Assuming an X-ray extent of 10\arcsec\ $\approx$ 2.4~pc (Section~\ref{sec:results}) is the radius of the wind bubble $R_{\rm b}$, we can use the \citet{Weaver77} model to estimate the bubble age $t$, the predicted temperature $T_{\rm b}$, and the X-ray luminosity $L_{\rm X}$. From Equation~21 of \cite{Weaver77},

\begin{equation}
t = \bigg( \frac{125}{154 \pi}\bigg)^{-1/3} L_{\rm w}^{-1/3} \rho^{1/3} R_{\rm b}^{5/3}
\label{eq:R_weaver}
\end{equation}

\noindent
which yields $t = 2.4\times10^{4}$~yr. This age is younger than the estimated age of the YSOs; e.g. the molecular gas outflows from \starcluster\ have an associated timescale of $\approx 6.5\times10^{4}~{\rm yr}$ \citep{Nayak19}. From $t$, the Weaver model predicts a bubble temperature $T_{\rm b}$ of (their Equation 37)

\begin{equation}
T_{\rm b} = 2.07\times10^{6} L_{36}^{8/35} n_{0}^{2/35}t_{6}^{-6/35}~{\rm K},
\label{eq:T_Weaver}
\end{equation}

\noindent
where $L_{36} \equiv L_{\rm w}/10^{36}~{\rm erg~s}^{-1}$, $t_{6} \equiv t / 10^{6}~{\rm yr}$, and $n_{0} = \rho / \mu m_{\rm p}$ is the ambient number density. Assuming a mean molecular weight of $\mu = 1.4$ for the background gas, we find $T_{\rm b} = 1.0\times10^{7}$~K, corresponding to $kT = 0.9$~keV. This value is statistically lower than our estimated $kT = 3.5^{+32.1}_{-2.2}$~keV from X-ray spectral modeling (see Table~\ref{table:spectralresults}). As the \citet{Weaver77} model does not take into account cooling at the bubble-shell interface, it over-estimates the mass-loading of the bubble interior, leading to an under-estimate of $T_{\rm b}$ \citep[see e.g. Eq. 45 of ][]{ElBadry19}, consistent with what we see here.

\begin{deluxetable*}{lcccccr}[!t]
\tablecolumns{7}
\tablewidth{0pt} \tablecaption{Characteristics of Star-Forming Regions with Detected Diffuse X-ray Emission\label{table:SFR Characteristics}} 
\tablehead{\colhead{Region} & \colhead{Age} & \colhead{Radius\tablenotemark{a}} & \colhead{Distance} & \colhead{$L_{\rm bol}$} & \colhead{$L_{\rm X}$} & \colhead{References}\\
\colhead{} & \colhead{(Myr)} & \colhead{(pc)} & \colhead{(kpc)} & \colhead{($L_{\sun}$)} & \colhead{(${\rm erg}\, {\rm s}^{-1})$} & \colhead{}}
\startdata
RCW 38 & 0.1$-$0.5 & 10 & 1.7 & $9.7 \times 10^5$ & $1.5\times10^{33}$ & 1, 2, 3, 4\\
N79 & 0.5 & 64 & 50 & $2.2 \times 10^6$ & $1.5 \times 10^{34}$ & this work, 5, 6\\
NGC 3603\tablenotemark{b,c} & 1 & 21 & 7.0 & $2.3 \times 10^7$ & $2.6 \times 10^{35}$ & 7, 8, 9, 10\\
Arches Cluster\tablenotemark{d,e} & 1$-$2 & 3.6 & 8.5 & ... & $1.6 \times 10^{34}$ & 11\\
30 Doradus\tablenotemark{b} & 1$-$2 & 100 & 50 & $7.8 \times 10^7$ & $4.5 \times 10^{36}$ & 12, 13\\
Carina Nebula & 2$-$3 & 20 & 2.3 & $2.5 \times 10^7$ & $1.7 \times 10^{35}$ & 10, 14,\\
Rosette Nebula & 2$-$4 & 16.5 & 1.4 & $2.3 \times 10^6$ & $6.0 \times 10^{32}$ & 15, 16, 17, 18\\
Westerlund 1\tablenotemark{f} & 3$-$4 & 25 & 4.0 & $2.6 \times 10^6$ & $3.0 \times 10^{34}$ & 9, 19, 20, 21\\
\enddata
\tablenotetext{a}{Radius defined by the H-alpha emission.}
\tablenotetext{b}{$L_{\rm X}$ corresponding to the soft-band (0.5$-$2.0 keV) diffuse emission.}
\tablenotetext{c}{Radius corresponding to the radio flux.}
\tablenotetext{d}{Due to extinction uncertainties toward the Galactic Center, this $L_{\rm bol}$ has substantial uncertainty \citep{Clark18}.}
\tablenotetext{e}{Radius corresponding to the maximum extent of the diffuse X-ray emission (3.6pc$\times$2.4pc).}
\tablenotetext{f}{Radius corresponding to the extent of the H~{\sc i} bubble.}
\tablerefs{(1) \cite{Rodgers60}; (2) \cite{Wolk06}; (3) \cite{Fukui16}; (4) \cite{Pandey24}; (5) \cite{Lopez14}; (6) \cite{Ochsendorf17}; (7) \cite{Balick80}; (8) \cite{Moffat02}; (9) \cite{Binder18}; (10) \cite{Townsley11}; (11) \cite{YZ02}; (12) \cite{Lopez11}; (13) \cite{Malumuth94}; (14) \cite{HarperClark09};  (15) \cite{Bruhweiler10}; (16) \cite{Celnik85}; (17) \cite{Cox90}; (18) \cite{Townsley03}; (19) \cite{Muno06}; (20) \cite{Kothes07}; (21) \cite{Dougherty10}}
\end{deluxetable*}

Finally, we combine the above derived quantities with X-ray emissivities calculated with \texttt{yt}'s X-ray emissivity\footnote{See https://hea-www.cfa.harvard.edu/$\sim$jzuhone/pyxsim/ \\ xray\_fields.html} calculator \citep{Turk11} to find an associated X-ray luminosity in the $0.5-7\, {\rm keV}$ band for the \cite{Weaver77} model of $L_{\rm X} = 3.4\times 10^{35}\, {\rm erg}\, {\rm s}^{-1}$. This value is about an order of magnitude larger than the observed $L_{\rm X}$ (see Table~\ref{table:spectralresults}). We can get another estimate of the X-ray luminosity from the \citet{Weaver77} model by following the derivation of Appendix B of \citet{ChuMacLow90}, which takes into account the temperature and density variation in the bubble's interior due to the evaporative mass flow in the \citet{Weaver77} model. Using their Equation B7 we infer an X-ray luminosity of $L_X = 1.0 \times 10^{35}\, {\rm erg}\, {\rm s}^{-1}$, which is still an order of magnitude larger than the observed value. 

These estimates, combined with the very young age inferred from the application of the \citet{Weaver77} model through Equation \ref{eq:R_weaver}, indicates that there is likely significant cooling at the wind bubble's interface. Such cooling would both reduce the efficiency of the bubble's expansion \citep{Rosen14, lancaster21a,lancaster21b}, explaining an older age, and reduce the mass-loading of the bubble from conductive evaporation \citep{ElBadry19}, explaining a lower $L_X$. Numerical simulations following the formation of individual massive stars have confirmed that these effects occur early \citep{Rosen2022}.

\subsection{Comparison to Other Young Star Clusters} \label{subsec:YSO}

Diffuse X-ray emission associated with gas shock-heated by stellar wind feedback has been detected in many other young star-forming regions, including NGC 3603 \citep{Moffat02}, the Arches Cluster \citep{YZ02}, RCW~38 \citep{Wolk02,fukushima23}, M17 \citep{Townsley03}, the Rosette Nebula \citep{Townsley03}, Westerlund 1 \citep{Muno06}, the Carina Nebula \citep{Townsley11}, and 30 Doradus \citep{Townsley06,Lopez11}. The characteristics of these star-forming regions are shown in Table~\ref{table:SFR Characteristics}, spanning two orders of magnitude in bolometric luminosity $L_{\rm bol}$ and four orders of magnitude in $L_{\rm X}$.

One region with a young age similar to \starcluster\ is the Milky Way H{\sc ii} region RCW~38 which is estimated to be 0.1$-$0.5 Myr old \citep{Wolk06,Fukui16}. RCW~38 has $\sim$80 times lower bolometric luminosity (with $L_{\rm bol} = 9.7\times10^{5}~L_{\sun}$), an order of magnitude lower X-ray luminosity (with $L_{\rm X} =(1.5\pm0.2)\times10^{33}~{\rm erg~s}^{-1}$), and a slightly higher hot gas temperature (with $kT = 4.5^{+1.2}_{-0.9} $~keV) than \starcluster\ \citep{Pandey24}. The diffuse X-ray emission of RCW~38 requires both thermal and non-thermal components to fit the spectra \citep{Wolk02,fukushima23,Pandey24}, whereas the weaker signal from \starcluster\ does not enable us to distinguish whether both components are necessary to model the data adequately. Both \starcluster\ and RCW~38 demonstrate that diffuse X-rays associated with stellar winds are produced early in MSC formation. 

\section{Conclusions} \label{sec:conc}

In this work, we present new Chandra observations totaling 98~ks toward the potential superstar cluster (SSC) \starcluster\ in N79-South. We detect $\sim$160 net, background-subtracted, broad-band X-ray counts from the 1\arcmin\ vicinity of \starcluster\ where five YSOs have been identified using JWST (see Figure~\ref{fig:3colorxray_panel}). We also serendipitously detect $\sim$12 net counts from another location in N79-South with three YSOs (see Figure~\ref{fig:3colorxray_S2panel}). The X-ray emission is extended $\sim$10\arcsec\ in radius, much greater than the Chandra on-axis PSF, which demonstrates that diffuse hot gas is produced by stellar-wind feedback in the earliest stages of formation.

We show that the X-ray emission around \starcluster\ is especially hard, dominated by photons above 1.2~keV, suggesting a high hot gas temperature, a large absorbing column in the region, presence of colliding-wind massive binaries, and/or contribution from a non-thermal/power-law component. The X-rays appear to be spatially anti-coincident with the $^{13}$CO dense gas (Figure~\ref{fig:CO13vXray}), suggesting that the hot gas is preferentially occupying the lower-density cavities and/or that X-rays are absorbed by the dense gas.

Comparison to stellar wind model predictions shows that the X-ray luminosity of \starcluster\ is about one order of magnitude below expected if the shock-heated gas is confined by a cool shell that heats up via thermal conduction and evaporates. This result suggest that, even this early in the MSC formation process, significant amounts of wind energy are being lost. Likely explanations are either turbulent mixing followed by radiative cooling \citep{Rosen14,lancaster21a,lancaster21b} or physical leakage of the gas \citep{HarperClark09}, with the former likely playing a larger role \citep{lancaster21c}.

\software{CIAO (v4.15; \citealt{CIAO2006}), XSPEC (v12.13.0c; \citealt{XSPEC})}

\begin{acknowledgements}

We are grateful for the helpful feedback from Marta Sewilo and the anonymous referee. Support for this work was provided by the National Aeronautics and Space Administration through Chandra Award Number GO0-21071X issued by the Chandra X-ray Center, which is operated by the Smithsonian Astrophysical Observatory for and on behalf of the National Aeronautics Space Administration under contract NAS8-03060. JR and LAL also acknowledge support through the Heising-Simons Foundation grant 2022-3533. LAL and LL gratefully acknowledges the support of the Simons Foundation. 

\end{acknowledgements}

\bibliographystyle{aasjournal}
\bibliography{bibliography}{}

\begin{thebibliography}{}
\expandafter\ifx\csname natexlab\endcsname\relax\def\natexlab#1{#1}\fi
\providecommand{\url}[1]{\href{#1}{#1}}
\providecommand{\dodoi}[1]{doi:~\href{http://doi.org/#1}{\nolinkurl{#1}}}
\providecommand{\doeprint}[1]{\href{http://ascl.net/#1}{\nolinkurl{http://ascl.net/#1}}}
\providecommand{\doarXiv}[1]{\href{https://arxiv.org/abs/#1}{\nolinkurl{https://arxiv.org/abs/#1}}}

\bibitem[{{Agertz} {et~al.}(2013){Agertz}, {Kravtsov}, {Leitner}, \&
  {Gnedin}}]{Agertz13}
{Agertz}, O., {Kravtsov}, A.~V., {Leitner}, S.~N., \& {Gnedin}, N.~Y. 2013,
  \apj, 770, 25, \dodoi{10.1088/0004-637X/770/1/25}

\bibitem[{{Arnaud}(1996)}]{XSPEC}
{Arnaud}, K.~A. 1996, in Astronomical Society of the Pacific Conference Series,
  Vol. 101, Astronomical Data Analysis Software and Systems V, ed. G.~H.
  {Jacoby} \& J.~{Barnes}, 17

\bibitem[{{Asplund} {et~al.}(2009){Asplund}, {Grevesse}, {Sauval}, \&
  {Scott}}]{Asplund09}
{Asplund}, M., {Grevesse}, N., {Sauval}, A.~J., \& {Scott}, P. 2009, \araa, 47,
  481, \dodoi{10.1146/annurev.astro.46.060407.145222}

\bibitem[{{Balick} {et~al.}(1980){Balick}, {Boeshaar}, \& {Gull}}]{Balick80}
{Balick}, B., {Boeshaar}, G.~O., \& {Gull}, T.~R. 1980, \apj, 242, 584,
  \dodoi{10.1086/158493}

\bibitem[{{Binder} \& {Povich}(2018)}]{Binder18}
{Binder}, B.~A., \& {Povich}, M.~S. 2018, \apj, 864, 136,
  \dodoi{10.3847/1538-4357/aad7b2}

\bibitem[{{Bruhweiler} {et~al.}(2010){Bruhweiler}, {Freire Ferrero}, {Bourdin},
  \& {Gull}}]{Bruhweiler10}
{Bruhweiler}, F.~C., {Freire Ferrero}, R., {Bourdin}, M.~O., \& {Gull}, T.~R.
  2010, \apj, 719, 1872, \dodoi{10.1088/0004-637X/719/2/1872}

\bibitem[{{Cant{\'o}} {et~al.}(2000){Cant{\'o}}, {Raga}, \&
  {Rodr{\'\i}guez}}]{Canto00}
{Cant{\'o}}, J., {Raga}, A.~C., \& {Rodr{\'\i}guez}, L.~F. 2000, \apj, 536,
  896, \dodoi{10.1086/308983}

\bibitem[{{Castor} {et~al.}(1975){Castor}, {McCray}, \& {Weaver}}]{Castor75}
{Castor}, J., {McCray}, R., \& {Weaver}, R. 1975, \apjl, 200, L107,
  \dodoi{10.1086/181908}

\bibitem[{{Celnik}(1985)}]{Celnik85}
{Celnik}, W.~E. 1985, \aap, 144, 171

\bibitem[{{Chevalier} \& {Clegg}(1985)}]{Chevalier85}
{Chevalier}, R.~A., \& {Clegg}, A.~W. 1985, \nat, 317, 44,
  \dodoi{10.1038/317044a0}

\bibitem[{{Choi} {et~al.}(2016){Choi}, {Dotter}, {Conroy}, {Cantiello},
  {Paxton}, \& {Johnson}}]{Choi16}
{Choi}, J., {Dotter}, A., {Conroy}, C., {et~al.} 2016, \apj, 823, 102,
  \dodoi{10.3847/0004-637X/823/2/102}

\bibitem[{{Chu} \& {Mac Low}(1990)}]{ChuMacLow90}
{Chu}, Y.-H., \& {Mac Low}, M.-M. 1990, \apj, 365, 510, \dodoi{10.1086/169505}

\bibitem[{{Clark} {et~al.}(2018){Clark}, {Lohr}, {Najarro}, {Dong}, \&
  {Martins}}]{Clark18}
{Clark}, J.~S., {Lohr}, M.~E., {Najarro}, F., {Dong}, H., \& {Martins}, F.
  2018, \aap, 617, A65, \dodoi{10.1051/0004-6361/201832826}

\bibitem[{{Cox} {et~al.}(1990){Cox}, {Deharveng}, \& {Leene}}]{Cox90}
{Cox}, P., {Deharveng}, L., \& {Leene}, A. 1990, \aap, 230, 181

\bibitem[{{Dotter}(2016)}]{Dotter16}
{Dotter}, A. 2016, \apjs, 222, 8, \dodoi{10.3847/0067-0049/222/1/8}

\bibitem[{{Dougherty} {et~al.}(2010){Dougherty}, {Clark}, {Negueruela},
  {Johnson}, \& {Chapman}}]{Dougherty10}
{Dougherty}, S.~M., {Clark}, J.~S., {Negueruela}, I., {Johnson}, T., \&
  {Chapman}, J.~M. 2010, \aap, 511, A58, \dodoi{10.1051/0004-6361/200913505}

\bibitem[{{Egorov} {et~al.}(2023){Egorov}, {Kreckel}, {Sandstrom}, {Leroy},
  {Glover}, {Groves}, {Kruijssen}, {Barnes}, {Belfiore}, {Bigiel}, {Blanc},
  {Boquien}, {Cao}, {Chastenet}, {Chevance}, {Congiu}, {Dale}, {Emsellem},
  {Grasha}, {Klessen}, {Larson}, {Liu}, {Murphy}, {Pan}, {Pessa}, {Pety},
  {Rosolowsky}, {Scheuermann}, {Schinnerer}, {Sutter}, {Thilker}, {Watkins}, \&
  {Williams}}]{egorov23}
{Egorov}, O.~V., {Kreckel}, K., {Sandstrom}, K.~M., {et~al.} 2023, \apjl, 944,
  L16, \dodoi{10.3847/2041-8213/acac92}

\bibitem[{{El-Badry} {et~al.}(2019){El-Badry}, {Ostriker}, {Kim}, {Quataert},
  \& {Weisz}}]{ElBadry19}
{El-Badry}, K., {Ostriker}, E.~C., {Kim}, C.-G., {Quataert}, E., \& {Weisz},
  D.~R. 2019, \mnras, 490, 1961, \dodoi{10.1093/mnras/stz2773}

\bibitem[{{Fruscione} {et~al.}(2006){Fruscione}, {McDowell}, {Allen},
  {Brickhouse}, {Burke}, {Davis}, {Durham}, {Elvis}, {Galle}, {Harris},
  {Huenemoerder}, {Houck}, {Ishibashi}, {Karovska}, {Nicastro}, {Noble},
  {Nowak}, {Primini}, {Siemiginowska}, {Smith}, \& {Wise}}]{CIAO2006}
{Fruscione}, A., {McDowell}, J.~C., {Allen}, G.~E., {et~al.} 2006, in Society
  of Photo-Optical Instrumentation Engineers (SPIE) Conference Series, Vol.
  6270, Society of Photo-Optical Instrumentation Engineers (SPIE) Conference
  Series, ed. D.~R. {Silva} \& R.~E. {Doxsey}, 62701V,
  \dodoi{10.1117/12.671760}

\bibitem[{{Fukui} {et~al.}(2016){Fukui}, {Torii}, {Ohama}, {Hasegawa},
  {Hattori}, {Sano}, {Ohashi}, {Fujii}, {Kuwahara}, {Mizuno}, {Dawson},
  {Yamamoto}, {Tachihara}, {Okuda}, {Onishi}, \& {Mizuno}}]{Fukui16}
{Fukui}, Y., {Torii}, K., {Ohama}, A., {et~al.} 2016, \apj, 820, 26,
  \dodoi{10.3847/0004-637X/820/1/26}

\bibitem[{{Fukushima} {et~al.}(2023){Fukushima}, {Ezoe}, \&
  {Odaka}}]{fukushima23}
{Fukushima}, A., {Ezoe}, Y., \& {Odaka}, H. 2023, \pasj, 75, 187,
  \dodoi{10.1093/pasj/psac100}

\bibitem[{{Gordon} {et~al.}(2008){Gordon}, {Engelbracht}, {Rieke}, {Misselt},
  {Smith}, \& {Kennicutt}}]{gordon08}
{Gordon}, K.~D., {Engelbracht}, C.~W., {Rieke}, G.~H., {et~al.} 2008, \apj,
  682, 336, \dodoi{10.1086/589567}

\bibitem[{{Harper-Clark} \& {Murray}(2009)}]{HarperClark09}
{Harper-Clark}, E., \& {Murray}, N. 2009, \apj, 693, 1696,
  \dodoi{10.1088/0004-637X/693/2/1696}

\bibitem[{{Henize}(1956)}]{Henize56}
{Henize}, K.~G. 1956, \apjs, 2, 315, \dodoi{10.1086/190025}

\bibitem[{{HI4PI Collaboration} {et~al.}(2016){HI4PI Collaboration}, {Ben
  Bekhti}, {Fl{\"o}er}, {Keller}, {Kerp}, {Lenz}, {Winkel}, {Bailin},
  {Calabretta}, {Dedes}, {Ford}, {Gibson}, {Haud}, {Janowiecki}, {Kalberla},
  {Lockman}, {McClure-Griffiths}, {Murphy}, {Nakanishi}, {Pisano}, \&
  {Staveley-Smith}}]{HI}
{HI4PI Collaboration}, {Ben Bekhti}, N., {Fl{\"o}er}, L., {et~al.} 2016, \aap,
  594, A116, \dodoi{10.1051/0004-6361/201629178}

\bibitem[{{Hunter} {et~al.}(1995){Hunter}, {Shaya}, {Holtzman}, {Light},
  {O'Neil}, \& {Lynds}}]{Hunter95}
{Hunter}, D.~A., {Shaya}, E.~J., {Holtzman}, J.~A., {et~al.} 1995, \apj, 448,
  179, \dodoi{10.1086/175950}

\bibitem[{{Kothes} \& {Dougherty}(2007)}]{Kothes07}
{Kothes}, R., \& {Dougherty}, S.~M. 2007, \aap, 468, 993,
  \dodoi{10.1051/0004-6361:20077309}

\bibitem[{{Kroupa}(2001)}]{Kroupa01}
{Kroupa}, P. 2001, \mnras, 322, 231, \dodoi{10.1046/j.1365-8711.2001.04022.x}

\bibitem[{{Krumholz} {et~al.}(2019){Krumholz}, {McKee}, \&
  {Bland-Hawthorn}}]{Krumholz19}
{Krumholz}, M.~R., {McKee}, C.~F., \& {Bland-Hawthorn}, J. 2019, \araa, 57,
  227, \dodoi{10.1146/annurev-astro-091918-104430}

\bibitem[{{Kurt} \& {Dufour}(1998)}]{kurt98}
{Kurt}, C.~M., \& {Dufour}, R.~J. 1998, in Revista Mexicana de Astronomia y
  Astrofisica Conference Series, Vol.~7, Revista Mexicana de Astronomia y
  Astrofisica Conference Series, ed. R.~J. {Dufour} \& S.~{Torres-Peimbert},
  202

\bibitem[{{Lancaster} {et~al.}(2021{\natexlab{a}}){Lancaster}, {Ostriker},
  {Kim}, \& {Kim}}]{lancaster21a}
{Lancaster}, L., {Ostriker}, E.~C., {Kim}, J.-G., \& {Kim}, C.-G.
  2021{\natexlab{a}}, \apj, 914, 89, \dodoi{10.3847/1538-4357/abf8ab}

\bibitem[{{Lancaster} {et~al.}(2021{\natexlab{b}}){Lancaster}, {Ostriker},
  {Kim}, \& {Kim}}]{lancaster21b}
---. 2021{\natexlab{b}}, \apj, 914, 90, \dodoi{10.3847/1538-4357/abf8ac}

\bibitem[{{Lancaster} {et~al.}(2021{\natexlab{c}}){Lancaster}, {Ostriker},
  {Kim}, \& {Kim}}]{lancaster21c}
---. 2021{\natexlab{c}}, \apjl, 922, L3, \dodoi{10.3847/2041-8213/ac3333}

\bibitem[{{Lopez} {et~al.}(2011){Lopez}, {Krumholz}, {Bolatto}, {Prochaska}, \&
  {Ramirez-Ruiz}}]{Lopez11}
{Lopez}, L.~A., {Krumholz}, M.~R., {Bolatto}, A.~D., {Prochaska}, J.~X., \&
  {Ramirez-Ruiz}, E. 2011, \apj, 731, 91, \dodoi{10.1088/0004-637X/731/2/91}

\bibitem[{{Lopez} {et~al.}(2014){Lopez}, {Krumholz}, {Bolatto}, {Prochaska},
  {Ramirez-Ruiz}, \& {Castro}}]{Lopez14}
{Lopez}, L.~A., {Krumholz}, M.~R., {Bolatto}, A.~D., {et~al.} 2014, \apj, 795,
  121, \dodoi{10.1088/0004-637X/795/2/121}

\bibitem[{{Madden} {et~al.}(2006){Madden}, {Galliano}, {Jones}, \&
  {Sauvage}}]{madden06}
{Madden}, S.~C., {Galliano}, F., {Jones}, A.~P., \& {Sauvage}, M. 2006, \aap,
  446, 877, \dodoi{10.1051/0004-6361:20053890}

\bibitem[{{Maggi} {et~al.}(2016){Maggi}, {Haberl}, {Kavanagh}, {Sasaki},
  {Bozzetto}, {Filipovi{\'c}}, {Vasilopoulos}, {Pietsch}, {Points}, {Chu},
  {Dickel}, {Ehle}, {Williams}, \& {Greiner}}]{maggi16}
{Maggi}, P., {Haberl}, F., {Kavanagh}, P.~J., {et~al.} 2016, \aap, 585, A162,
  \dodoi{10.1051/0004-6361/201526932}

\bibitem[{{Malumuth} \& {Heap}(1994)}]{Malumuth94}
{Malumuth}, E.~M., \& {Heap}, S.~R. 1994, \aj, 107, 1054,
  \dodoi{10.1086/116917}

\bibitem[{{Meixner} {et~al.}(2006){Meixner}, {Gordon}, {Indebetouw}, {Hora},
  {Whitney}, {Blum}, {Reach}, {Bernard}, {Meade}, {Babler}, {Engelbracht},
  {For}, {Misselt}, {Vijh}, {Leitherer}, {Cohen}, {Churchwell}, {Boulanger},
  {Frogel}, {Fukui}, {Gallagher}, {Gorjian}, {Harris}, {Kelly}, {Kawamura},
  {Kim}, {Latter}, {Madden}, {Markwick-Kemper}, {Mizuno}, {Mizuno}, {Mould},
  {Nota}, {Oey}, {Olsen}, {Onishi}, {Paladini}, {Panagia}, {Perez-Gonzalez},
  {Shibai}, {Sato}, {Smith}, {Staveley-Smith}, {Tielens}, {Ueta}, {van Dyk},
  {Volk}, {Werner}, \& {Zaritsky}}]{Meixner06}
{Meixner}, M., {Gordon}, K.~D., {Indebetouw}, R., {et~al.} 2006, \aj, 132,
  2268, \dodoi{10.1086/508185}

\bibitem[{{Moffat} {et~al.}(2002){Moffat}, {Corcoran}, {Stevens}, {Skalkowski},
  {Marchenko}, {M{\"u}cke}, {Ptak}, {Koribalski}, {Brenneman}, {Mushotzky},
  {Pittard}, {Pollock}, \& {Brandner}}]{Moffat02}
{Moffat}, A.~F.~J., {Corcoran}, M.~F., {Stevens}, I.~R., {et~al.} 2002, \apj,
  573, 191, \dodoi{10.1086/340491}

\bibitem[{{Montillaud} {et~al.}(2013){Montillaud}, {Joblin}, \&
  {Toublanc}}]{montillaud13}
{Montillaud}, J., {Joblin}, C., \& {Toublanc}, D. 2013, \aap, 552, A15,
  \dodoi{10.1051/0004-6361/201220757}

\bibitem[{{Muno} {et~al.}(2006){Muno}, {Law}, {Clark}, {Dougherty}, {de Grijs},
  {Portegies Zwart}, \& {Yusef-Zadeh}}]{Muno06}
{Muno}, M.~P., {Law}, C., {Clark}, J.~S., {et~al.} 2006, \apj, 650, 203,
  \dodoi{10.1086/507175}

\bibitem[{{Nayak} {et~al.}(2019){Nayak}, {Meixner}, {Sewi{\l}o}, {Ochsendorf},
  {Bolatto}, {Indebetouw}, {Kawamura}, {Onishi}, \& {Fukui}}]{Nayak19}
{Nayak}, O., {Meixner}, M., {Sewi{\l}o}, M., {et~al.} 2019, \apj, 877, 135,
  \dodoi{10.3847/1538-4357/ab1b38}

\bibitem[{{Nayak} {et~al.}(2024){Nayak}, {Hirschauer}, {Kavanagh}, {Meixner},
  {Chu}, {Habel}, {Jones}, {Lenki{\'c}}, {Nally}, {Reiter}, {Robberto}, \&
  {Sargent}}]{Nayak24}
{Nayak}, O., {Hirschauer}, A.~S., {Kavanagh}, P.~J., {et~al.} 2024, \apj, 963,
  94, \dodoi{10.3847/1538-4357/ad18bc}

\bibitem[{{Ochsendorf} {et~al.}(2017){Ochsendorf}, {Zinnecker}, {Nayak},
  {Bally}, {Meixner}, {Jones}, {Indebetouw}, \& {Rahman}}]{Ochsendorf17}
{Ochsendorf}, B.~B., {Zinnecker}, H., {Nayak}, O., {et~al.} 2017, Nature
  Astronomy, 1, 784, \dodoi{10.1038/s41550-017-0268-0}

\bibitem[{{Pandey} {et~al.}(2024){Pandey}, {Lopez}, {Rosen}, {Thompson},
  {Linden}, {Blackstone}, \& {Lancaster}}]{Pandey24}
{Pandey}, P., {Lopez}, L.~A., {Rosen}, A.~L., {et~al.} 2024, arXiv e-prints,
  arXiv:2404.19001, \dodoi{10.48550/arXiv.2404.19001}

\bibitem[{{Points} {et~al.}(2024){Points}, {Long}, {Blair}, {Williams}, {Chu},
  {Winkler}, {White}, {Rest}, {Li}, \& {Valdes}}]{Points24}
{Points}, S.~D., {Long}, K.~S., {Blair}, W.~P., {et~al.} 2024, \apj, 974, 70,
  \dodoi{10.3847/1538-4357/ad6766}

\bibitem[{{Rauw} \& {Naz{\'e}}(2016)}]{Rauw16}
{Rauw}, G., \& {Naz{\'e}}, Y. 2016, Advances in Space Research, 58, 761,
  \dodoi{10.1016/j.asr.2015.09.026}

\bibitem[{{Rodgers} {et~al.}(1960){Rodgers}, {Campbell}, \&
  {Whiteoak}}]{Rodgers60}
{Rodgers}, A.~W., {Campbell}, C.~T., \& {Whiteoak}, J.~B. 1960, \mnras, 121,
  103, \dodoi{10.1093/mnras/121.1.103}

\bibitem[{{Rosen}(2022)}]{Rosen2022}
{Rosen}, A.~L. 2022, \apj, 941, 202, \dodoi{10.3847/1538-4357/ac9f3d}

\bibitem[{{Rosen} {et~al.}(2014){Rosen}, {Lopez}, {Krumholz}, \&
  {Ramirez-Ruiz}}]{Rosen14}
{Rosen}, A.~L., {Lopez}, L.~A., {Krumholz}, M.~R., \& {Ramirez-Ruiz}, E. 2014,
  \mnras, 442, 2701, \dodoi{10.1093/mnras/stu1037}

\bibitem[{{Seale} {et~al.}(2014){Seale}, {Meixner}, {Sewi{\l}o}, {Babler},
  {Engelbracht}, {Gordon}, {Hony}, {Misselt}, {Montiel}, {Okumura}, {Panuzzo},
  {Roman-Duval}, {Sauvage}, {Boyer}, {Chen}, {Indebetouw}, {Matsuura},
  {Oliveira}, {Srinivasan}, {van Loon}, {Whitney}, \& {Woods}}]{Seale14}
{Seale}, J.~P., {Meixner}, M., {Sewi{\l}o}, M., {et~al.} 2014, \aj, 148, 124,
  \dodoi{10.1088/0004-6256/148/6/124}

\bibitem[{{Smith}(2006)}]{Smith06}
{Smith}, N. 2006, \mnras, 367, 763, \dodoi{10.1111/j.1365-2966.2006.10007.x}

\bibitem[{{Smith} \& {MCELS Team}(1998)}]{Smith98}
{Smith}, R.~C., \& {MCELS Team}. 1998, \pasa, 15, 163, \dodoi{10.1071/AS98163}

\bibitem[{{Stevens} \& {Hartwell}(2003)}]{Stevens03}
{Stevens}, I.~R., \& {Hartwell}, J.~M. 2003, \mnras, 339, 280,
  \dodoi{10.1046/j.1365-8711.2003.06184.x}

\bibitem[{{Townsley} {et~al.}(2011){Townsley}, {Broos}, {Chu}, {Gruendl},
  {Oey}, \& {Pittard}}]{Townsley11}
{Townsley}, L.~K., {Broos}, P.~S., {Chu}, Y.-H., {et~al.} 2011, \apjs, 194, 16,
  \dodoi{10.1088/0067-0049/194/1/16}

\bibitem[{{Townsley} {et~al.}(2006){Townsley}, {Broos}, {Feigelson}, {Brandl},
  {Chu}, {Garmire}, \& {Pavlov}}]{Townsley06}
{Townsley}, L.~K., {Broos}, P.~S., {Feigelson}, E.~D., {et~al.} 2006, \aj, 131,
  2140, \dodoi{10.1086/500532}

\bibitem[{{Townsley} {et~al.}(2024){Townsley}, {Broos}, \&
  {Povich}}]{Townsley24}
{Townsley}, L.~K., {Broos}, P.~S., \& {Povich}, M.~S. 2024, arXiv e-prints,
  arXiv:2403.16944, \dodoi{10.48550/arXiv.2403.16944}

\bibitem[{{Townsley} {et~al.}(2003){Townsley}, {Feigelson}, {Montmerle},
  {Broos}, {Chu}, \& {Garmire}}]{Townsley03}
{Townsley}, L.~K., {Feigelson}, E.~D., {Montmerle}, T., {et~al.} 2003, \apj,
  593, 874, \dodoi{10.1086/376692}

\bibitem[{{Turk} {et~al.}(2011){Turk}, {Smith}, {Oishi}, {Skory}, {Skillman},
  {Abel}, \& {Norman}}]{Turk11}
{Turk}, M.~J., {Smith}, B.~D., {Oishi}, J.~S., {et~al.} 2011, The Astrophysical
  Journal Supplement Series, 192, 9, \dodoi{10.1088/0067-0049/192/1/9}

\bibitem[{{Vink} {et~al.}(2000){Vink}, {de Koter}, \& {Lamers}}]{Vink00}
{Vink}, J.~S., {de Koter}, A., \& {Lamers}, H.~J.~G.~L.~M. 2000, \aap, 362,
  295, \dodoi{10.48550/arXiv.astro-ph/0008183}

\bibitem[{{Vink} {et~al.}(2001){Vink}, {de Koter}, \& {Lamers}}]{Vink01}
---. 2001, \aap, 369, 574, \dodoi{10.1051/0004-6361:20010127}

\bibitem[{{Weaver} {et~al.}(1977){Weaver}, {McCray}, {Castor}, {Shapiro}, \&
  {Moore}}]{Weaver77}
{Weaver}, R., {McCray}, R., {Castor}, J., {Shapiro}, P., \& {Moore}, R. 1977,
  \apj, 218, 377, \dodoi{10.1086/155692}

\bibitem[{{Willingale} {et~al.}(2013){Willingale}, {Starling}, {Beardmore},
  {Tanvir}, \& {O'Brien}}]{Willingale13}
{Willingale}, R., {Starling}, R.~L.~C., {Beardmore}, A.~P., {Tanvir}, N.~R., \&
  {O'Brien}, P.~T. 2013, \mnras, 431, 394, \dodoi{10.1093/mnras/stt175}

\bibitem[{{Wilms} {et~al.}(2000){Wilms}, {Allen}, \& {McCray}}]{wilms00}
{Wilms}, J., {Allen}, A., \& {McCray}, R. 2000, \apj, 542, 914,
  \dodoi{10.1086/317016}

\bibitem[{{Wolk} {et~al.}(2002){Wolk}, {Bourke}, {Smith}, {Spitzbart}, \&
  {Alves}}]{Wolk02}
{Wolk}, S.~J., {Bourke}, T.~L., {Smith}, R.~K., {Spitzbart}, B., \& {Alves}, J.
  2002, \apjl, 580, L161, \dodoi{10.1086/345611}

\bibitem[{{Wolk} {et~al.}(2006){Wolk}, {Spitzbart}, {Bourke}, \&
  {Alves}}]{Wolk06}
{Wolk}, S.~J., {Spitzbart}, B.~D., {Bourke}, T.~L., \& {Alves}, J. 2006, \aj,
  132, 1100, \dodoi{10.1086/505704}

\bibitem[{{Wong} {et~al.}(2011){Wong}, {Hughes}, {Ott}, {Muller}, {Pineda},
  {Bernard}, {Chu}, {Fukui}, {Gruendl}, {Henkel}, {Kawamura}, {Klein},
  {Looney}, {Maddison}, {Mizuno}, {Paradis}, {Seale}, \& {Welty}}]{Wong11}
{Wong}, T., {Hughes}, A., {Ott}, J., {et~al.} 2011, \apjs, 197, 16,
  \dodoi{10.1088/0067-0049/197/2/16}

\bibitem[{{Yusef-Zadeh} {et~al.}(2002){Yusef-Zadeh}, {Law}, {Wardle}, {Wang},
  {Fruscione}, {Lang}, \& {Cotera}}]{YZ02}
{Yusef-Zadeh}, F., {Law}, C., {Wardle}, M., {et~al.} 2002, \apj, 570, 665,
  \dodoi{10.1086/340058}

\end{thebibliography}

\end{document}